\documentclass[twocolumn]{aastex701}
\usepackage{siunitx}
\usepackage{textcomp}
\usepackage{booktabs}

\begin{document}

\title{Observation Timelines for the Potential Lunar Impact of Asteroid 2024~YR\textsubscript{4}}

\author{Yifan He}
\affiliation{Tsinghua University, Beijing 100084, China}
\email[]{yf-he22@mails.tsinghua.edu.cn}  

\author{Yixuan Wu}
\affiliation{Tsinghua University, Beijing 100084, China}
\email[]{}

\author{Yifei Jiao} 
\affiliation{Tsinghua University, Beijing 100084, China}
\affiliation{Department of Earth and Planetary Science, University of California, Santa Cruz, CA 95064, USA}
\email[show]{jiaoyf.thu@gmail.com}

\author{Wen-Yue Dai} 
\affiliation{Tsinghua University, Beijing 100084, China}
\affiliation{National Astronomical Observatory of Japan, 2-21-1 Osawa, Mitaka, Tokyo 181-8588, Japan}
\email[]{}

\author{Xin Liu}
\affiliation{Nanjing University, Nanjing 210023, Jiangsu, China}
\email[]{}

\author{Bin Cheng}
\affiliation{Tsinghua University, Beijing 100084, China}
\email[show]{bincheng@tsinghua.edu.cn}

\author{Hexi Baoyin}
\affiliation{Tsinghua University, Beijing 100084, China}
\affiliation{Inner Mongolia University of Technology, Hohhot 010051, China}
\email[]{baoyin@tsinghua.edu.cn}

\begin{abstract}

The near-Earth asteroid 2024~YR\textsubscript{4}—a $\sim$60\,m rocky object that was once considered a potential Earth impactor—has since been ruled out for Earth but retained a $\sim$4.3\% probability of striking the Moon in 2032. Such an impact, with equivalent kinetic energy of $\sim$6.5\,Mt TNT, is expected to produce a $\sim$1\,km crater on the Moon, and will be the most energetic lunar impact event ever recorded in human history. Despite the associated risk, this scenario offers a rare and valuable scientific opportunity.
Using a hybrid framework combining Monte Carlo orbital propagation, smoothed particle hydrodynamics (SPH) impact modeling, and N‐body ejecta dynamics, we evaluate the physical outcomes and propose the observation timelines of this rare event. 
Our results suggest an optical flash of visual magnitude from $-$2.5 to $-$3 lasting several minutes directly after the impact, followed by hours of infrared afterglow from $\sim$2000\,K molten rock cooling to a few hundreds K. The associated seismic energy release would lead to a global-scale lunar reverberation (magnitude $\sim$5.0) that can be detectable by any modern seismometers. Furthermore, the impact would throw out $\sim10^8$\,kg debris to escape the lunar gravity, with a small fraction reaching Earth to produce a lunar meteor outburst within 100 years.  
Finally, we integrate these results into a coordinated observation timeline, identifying the best detection windows for ground‐based telescopes, lunar orbiters and surface stations.
\end{abstract}



\section{INTRODUCTION}

Asteroid 2024~YR\textsubscript{4} was discovered in late 2024 as a $\sim$60 m Apollo-class near-Earth asteroid \citep{bolin2025discoveryYR4,rivkin2025jwst} whose orbit crosses those of both Earth and the Moon. 
Refined orbital solutions in mid-2025 \citep{NASA2025YR4} indicate a 4.3\% probability that 2024~YR\textsubscript{4} will strike the Moon on 2032 December 22 at a velocity of $\sim$14.1~km\,s$^{-1}$. 
Should this occur, 2024~YR\textsubscript{4} would deliver a kinetic energy of $\sim$$3\times10^{16}$ J—about 6.5 megatons of TNT—excavating a $\sim$1 km-scale crater with a depth of $\sim 150\ \rm m$. 
Such an event would constitute the most energetic lunar impact in several millennia and an unprecedented natural experiment in impact physics.

Only far smaller lunar impacts have ever been directly observed. For example, the 2013 September 11 event \citep{madiedo2014large}, produced by a $\sim$400 kg meteoroid, generated a brief 3rd-magnitude flash and a crater only 40 meters wide. 
The 2024~YR\textsubscript{4} impact, by contrast, would release energy 6 orders of magnitude greater—comparable to a medium nuclear detonation—and would trigger a suite of observable phenomena: an intense optical and infrared flash, a km-scale melt pool, a global moonquake, and a short-lived cloud of escaping ejecta. 
Historical Apollo seismometers recorded even the $\sim$1 kT TNT-equivalent impact of the Apollo 13 S-IVB stage at global distances \citep{gudkova2011large}, implying that a 2024~YR\textsubscript{4}-sized collision would produce seismic signals readily detectable across the Moon. 
Moreover, the fastest ejecta could exceed the lunar escape speed, temporarily populating cislunar space and, within days, producing a meteoroid flux toward Earth—raising both scientific interests and hazard considerations for artificial satellites, as discussed in \citet{2025ApJ...990L..20W}.

Here we present a comprehensive analysis of the dynamical, thermal, and seismic consequences of a potential 2024~YR\textsubscript{4} impact and evaluate their observability across multiple domains. 
Our study combines high-precision Monte Carlo orbital propagation \citep{liu2025collision} with smoothed particle hydrodynamics (SPH) impact simulations \citep{jiao2024sph} and N-body modeling of high-velocity ejecta \citep{rein2012rebound}. 
From these results we derive quantitative predictions for the optical flash, crater formation, ejecta dispersion, thermal afterglow, and seismic wave amplitudes, linking each to specific observational platforms and timescales. 
We further integrate these outcomes into a coordinated observation framework encompassing ground-based telescopes, lunar orbiters, surface seismometers, and space-based assets, summarized in a unified timeline of detectability.  
Our work emphasizes the full chain of observable effects—from the sub-second optical flash to the multi-day infrared emission and global seismic response—and outlines how simultaneous monitoring across these domains could yield the most complete dataset ever obtained for a large lunar impact.  
If the predicted collision indeed come up, 2024~YR\textsubscript{4} will offer a once-in-a-generation opportunity to observe, in real time, the interplay between impact dynamics, lunar geology, and near-Earth space environment, providing crucial benchmarks for planetary-defense modeling and future lunar science.


\setcounter{footnote}{0} 

\section{METHODOLOGY} \label{sec:style}

\subsection{Monte Carlo Orbit Sampling and Impact Geometry}
To better constrain the impact corridor and simulation parameters, we generated $10^4$ orbit “clones” by sampling the $6\times6$ covariance of the nominal JPL/MPC solution \citep{JPL2025YR4,MPC2024YR4} (mid-2025) for 2024~YR\textsubscript{4}. 

Each clone’s Cartesian state at epoch was drawn from the covariance ellipsoid and numerically propagated to 2032 December using a high-precision N-body integrator with relativistic terms, DE441 ephemerides for all major bodies (Sun, eight planets, Pluto, Earth–Moon resolved; lunar gravity consistent with DE441), and an adaptive Radau/Prince–Dormand scheme with absolute tolerance $10^{-12}$ and base step $\sim$0.1\,d. 
Nongravitational forces were neglected over the 7–8\,yr horizon \citep{liu2025collision}.
A “lunar impact” is flagged when a trajectory crossed the lunar radius plus gravitational focusing. 

\begin{figure*}[ht!]
    \centering
    \includegraphics[width=0.9\linewidth]{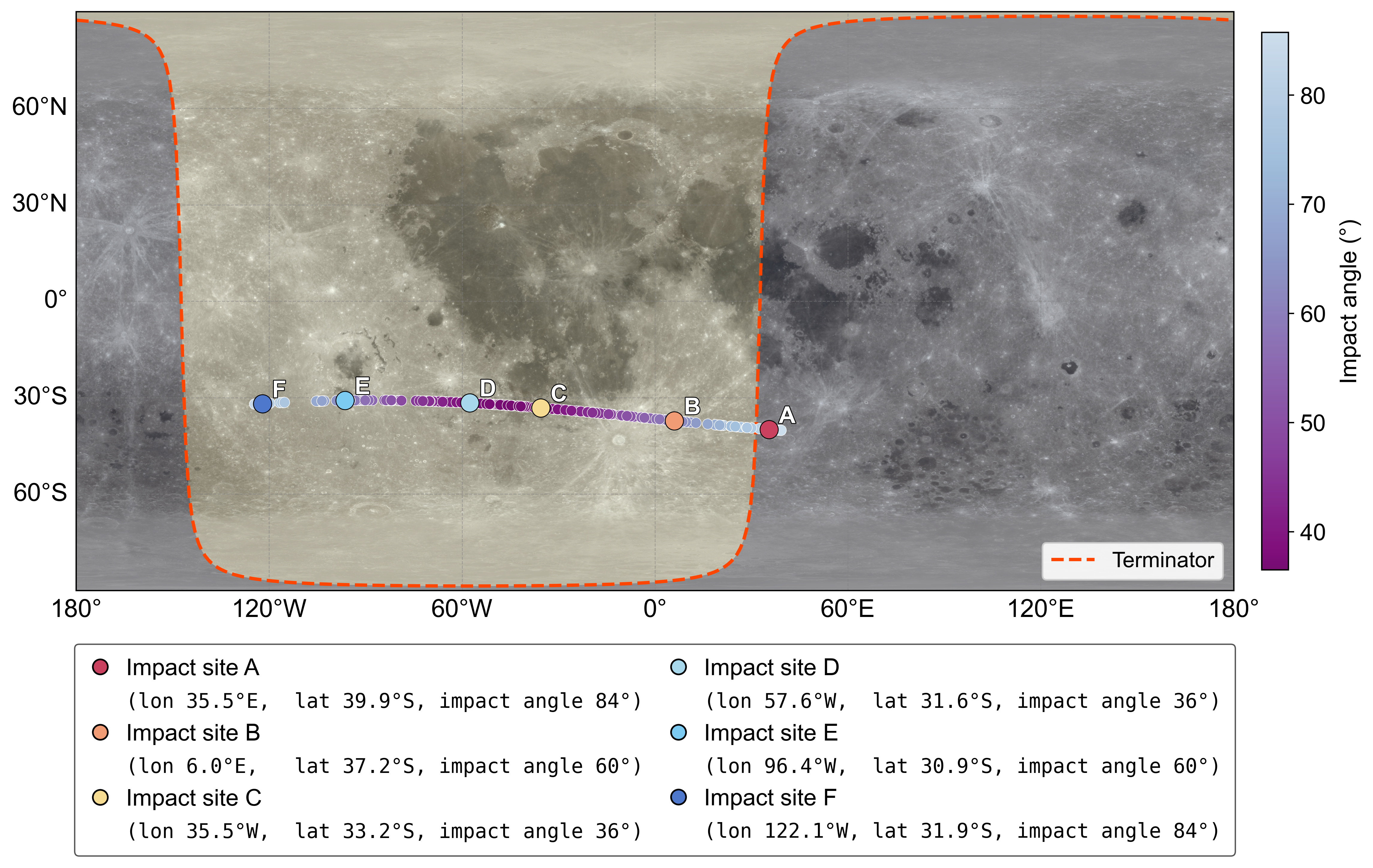} 
\caption{Map of the Moon’s entire surface showing the 4.3\% 2024~YR\textsubscript{4} impact corridor (with impact angle) and the dawn/dusk terminator (orange) on 22 December 2032 at 15:19 UTC. Six representative impact sites are highlighted, with their coordinates and impact angles listed in the bottom legend.}\label{fig:corridor}
\end{figure*}

From this single Monte Carlo campaign, we directly obtain the key impact statistics and geometry: $426/10{,}000$ clones intersect the Moon on 2032-12-22, implying $P_{\rm imp}=4.26\%$.
Impact times cluster within 2 hours of 15:18\,UTC. 
All impacts occur at $v_{\rm imp}=14.1$\,km\,s$^{-1}$ with $<0.01\%$ dispersion; incidence angles span $36^\circ-88^\circ$ (from vertical). 
The resulting impact corridor is shown in Figure~\ref{fig:corridor}. 
These consolidated outcomes are used directly to select angles and speed for impact modeling below.

\subsection{SPH Impact Simulation}

We simulate the impact process with SPH simulations for 500 s from the collision instant \citep{jiao2024sph, jiao2024asteroid}. 
Because the size of 2024~YR\textsubscript{4} (tens of meters) is far less than the Moon, the impact affects only a local region of the surface. 
Accordingly, the computational domain is taken as a hemispherical region of 3 km diameter centered on the impact point where $\sim6\times10^{6}$ particles are adopted, with lunar gravity approximated as a constant $g_{\rm Moon}=1.625$ m s$^{-2}$. 

The projectile is modeled as a 60 m rocky sphere (S-type proxy) at $v_{\rm imp}=14.1$ km s$^{-1}$ impacting into a local half-space of lunar regolith. 
Three representative incidences ($36^\circ$, $60^\circ$, $84^\circ$) are consistent with the Monte Carlo geometry. 
The material response is modeled using the Lund strength model, the Tillotson equation of state, and the $p$–$\alpha$ porosity compaction model. 
For the lunar surface, the shallow regolith layer is treated as a fully damaged, loosely consolidated granular medium with finite cohesion \citep{collins2004modeling, jutzi2015sph}. 
The basaltic material parameters are adopted for the equation of state \citep{melosh1989impact}, and the porosity model parameters follow the analysis of \citet{jutzi2008numerical, jutzi2009numerical}. 
The projectile (2024~YR\textsubscript{4}) is described by a von Mises strength model with a yield strength of 100 MPa \citep{cheng2024structural} and the Tillotson EOS for basalt. 

All material parameters applied in our simulations are summarized in Table~\ref{tab:eos_strength}.
The outcomes of our simulations are the transient/final crater dimensions, ejecta speed/angle distributions (including the high-velocity tail), and diagnostic fields (pressure, damage). 

\begingroup
\renewcommand{\arraystretch}{0.80}
\begin{deluxetable*}{lccccccccccccc c}
\setlength{\tabcolsep}{3pt}
\tabletypesize{\scriptsize}
\tablewidth{0pt}
\tablecaption{SPH material parameters adopted for the 2024~YR\textsubscript{4} lunar impact simulations\label{tab:eos_strength}}
\tablehead{
\colhead{Material} &
\colhead{$\rho_{0}$} &
\colhead{$A_{t}$} &
\colhead{$B_{t}$} &
\colhead{$a_{t}$} &
\colhead{$b_{t}$} &
\colhead{$\alpha_{t}$} &
\colhead{$\beta_{t}$} &
\colhead{$E_{0}$} &
\colhead{$E_{\rm iv}$} &
\colhead{$E_{\rm cv}$} &
\colhead{$Y_{i0}$} &
\colhead{$Y_{d0}$} &
\colhead{$Y_{m}$} &
\colhead{$\mu$} \\
\colhead{} &
\colhead{(kg\,m$^{-3}$)} &
\colhead{(GPa)} &
\colhead{(GPa)} &
\colhead{} &
\colhead{} &
\colhead{} &
\colhead{} &
\colhead{(MJ\,kg$^{-1}$)} &
\colhead{(MJ\,kg$^{-1}$)} &
\colhead{(MJ\,kg$^{-1}$)} &
\colhead{(Pa)} &
\colhead{(Pa)} &
\colhead{(Pa)} &
\colhead{}
}
\startdata
Lunar regolith (shallow) & 3400 & 26.7 & 26.7 & 0.5 & 1.5 & 5.0 & 5.0 & 487  & 18.2 & 4.72 & -- & $1.0\times10^{3}$ & $1.0\times10^{9}$ & 0.6 \\
Asteroid 2024~YR\textsubscript{4}       & 2700 & 26.7 & 26.7 & 0.5 & 1.5 & 5.0 & 5.0 & 487  & 18.2 & 4.72 & $1.0\times10^{8}$ & 0         & $3.5\times10^{9}$ & 0.8 \\
\enddata
\tablecomments{%
$\rho_{0}$ initial density; $A_{t},B_{t}$ Tillotson coefficients; $a_{t},b_{t},\alpha_{t},\beta_{t}$ dimensionless EOS parameters;
$E_{0},E_{\rm iv},E_{\rm cv}$ specific energies; $Y_{i0}$, $Y_{d0}$, $Y_{m}$ strength terms; $\mu$ internal friction.}
\end{deluxetable*}
\endgroup

\subsection{High-Velocity Ejecta Evolution}

To model the orbital fate of escaping fragments, we construct six launch cases, corresponding to two representative points along the lunar impact corridor for each of the three incidence angles ($36^\circ$, $60^\circ$, $84^\circ$). For each case, test particles are released from the lunar surface within a conical launch geometry of half-angle $45^\circ$, following the ejecta pattern in our impact simulation. The truncated power-law distribution of a particle swarm with a total number of $N_0$ is derived from the SPH outputs and approximated by

\begin{equation}
N(>v)\propto v^{-\gamma}.
\end{equation}

For each launch site, we perform Monte Carlo realizations of 20,000 particles, drawing velocity magnitudes from the above distribution and azimuths randomly within the cone signed in Figure~\ref{fig:corridor}.

The particles are mapped into Earth–Moon barycentric inertial states at the instant of launch and propagated using the N-body simulation code REBOUND \citep{rein2012rebound}. During the 100-year orbital evolution, particles are advanced in a full heliocentric model including the Moon and all eight planets. 
Non-gravitational perturbations are neglected, as the solar radiation pressure is less than $10^{-3}$ of Earth's gravity at the lunar distance for the minimum fragment size considered ($D \sim 1$~mm) \citep{burns1979radiation,wiegert2025potential,yu2017ejecta}.
Each particle is treated as massless in the N-body integration, but representative sizes and masses are assigned post-facto according to a cumulative power-law size–frequency distribution of escaping fragments. The size distribution is written as
\begin{equation}
N(>D) = N_{0,D} D^{-\beta}
\label{eq:2}
\end{equation}
where the index $\beta$ is about 3.4--3.6 according to SPH simulations \citep{jiao2024asteroid}. 
We adopt an upper cutoff diameter of $D_{\rm max} \approx 2$~m, corresponding to $\sim 0.035$ times the diameter of 2024~YR\textsubscript{4} \citep{wu2025detectability}.
$N_{0,D}$ is taken to match the total mass of escape ejecta in each case.
This provides a quantitative mapping between the dynamical test particles and the physical ejecta population. We record orbital states and collisions with large bodies of all ejecta. This setup thus links SPH-derived crater-scale ejection kinematics to the longer-term dispersal and impact fates of 2024~YR\textsubscript{4} ejecta.

\section{RESULTS} \label{sec:floats}

\subsection{SPH Impact Simulation Results} \label{sec:3.1}

\begin{figure*}[ht!]
    \centering
    \includegraphics[width=\linewidth]{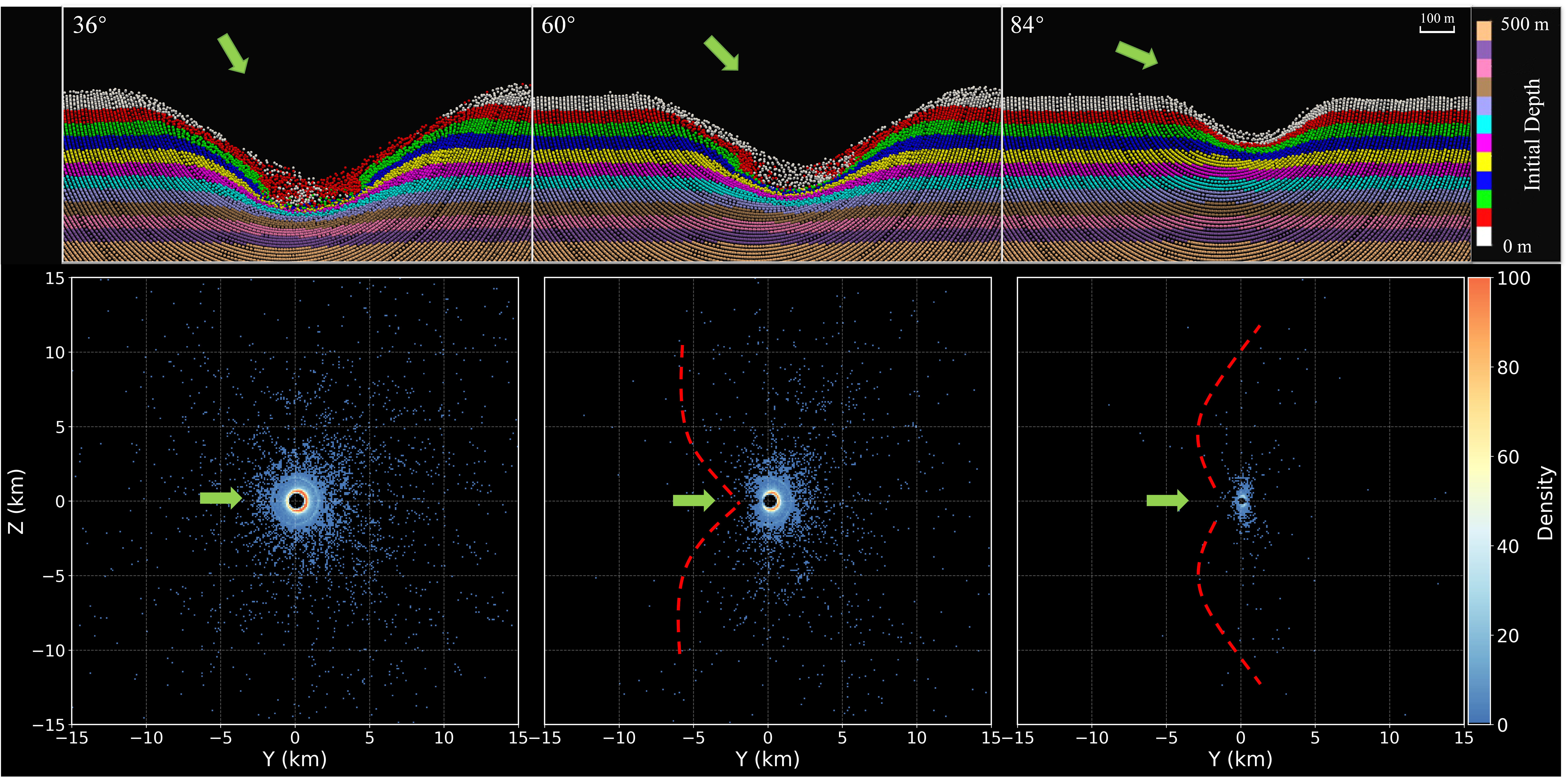} 
\caption{SPH outcomes for three incidence angles ($36^\circ$, $60^\circ$, $84^\circ$). Top row: crater cross‐sections at late time ($t\sim$200 s) showing size and depth differences. Bottom row: plan‐view ejecta patterns on the lunar surface, illustrating transition from near‐radial symmetry at $36^\circ$, to asymmetric butterfly rays at $60^\circ$, then to sparse wing-shaped ejecta landing points at $84^\circ$. The central black area marks the impact crater, and the region to the left of the red dashed line denotes the zone of avoidance (ZoA) with no ejecta relatively.}\label{fig:impact}
\end{figure*}

Shown in Figure~\ref{fig:impact}, in every case, 2024~YR\textsubscript{4} produces a simple bowl-shaped crater with a km-scale diameter, consistent with analytic scaling by \citet{collins2011size} and \citet{cheng2018collision}. 
At $36^\circ$ incidence, the crater stabilizes at $\sim$1.4 km diameter and $\sim$260 m depth, forming a circular bowl. The ejecta rays are radially distributed with no pronounced asymmetry. At $60^\circ$, the crater reaches $\sim$1.0 km diameter and $\sim$150 m depth ($d/D\simeq0.15$). The ejecta blanket displays a butterfly-shaped asymmetry, with rays extending tens of crater radii preferentially downrange. At the most oblique ($84^\circ$) incidence, the crater is smaller ($\sim0.6$ km) and shallower (120 m); its horizontal section is elliptic with axis ratio $\sim0.83$, and ejecta are sparse, confined to two lateral sectors with no coherent ray texture.These morphologies match established oblique impact systematics \citep{luo2022ejecta}.

The total mass of high-velocity ejecta escaping the Moon (exceeding 2.37 km\,s$^{-1}$) varies strongly with impact angle either. At the $36^\circ$ incidence (i.e., Craters C and D), the impact produces the most energetic ejection, yielding a total escaping mass of $\sim2.44\times10^{8}$\,kg. The $60^\circ$ case (i.e., Crater B and E) ejects about $1.94\times10^{8}$\,kg above the lunar escape speed, while the most oblique $84^\circ$ impacts (i.e., Craters A and F) generate only $\sim3.53\times10^{7}\,$kg of escaping fragments. The impact simulations were further used to fit the velocity–frequency power-law slopes of the escaping ejecta, yielding $\gamma\simeq1.01$, $0.87$, and $0.38$ for $36^\circ$, $60^\circ$, and $84^\circ$, respectively.

\subsection{Impact Flash and Thermal Afterglow}

The luminous output of a hypervelocity lunar impact is estimated by scaling the kinetic energy with a luminous efficiency $\eta$ \citep{merisio2023present}:
\begin{equation}
E_{\rm vis} = \eta\,E_{\rm imp}.
\end{equation}
Laboratory and lunar flash monitoring suggest $\eta \sim 10^{-3}$–$10^{-2}$ \citep{madiedo2014large, ait2015first}; here we explore the higher end ($\eta=10^{-2}$) to assess the bright-flash regime, justified by the relatively large size and rocky composition of 2024~YR\textsubscript{4}. For $E_{\rm imp}\sim3\times10^{16}$ J, this gives $E_{\rm vis} \simeq 3\times10^{14}\,{\rm J}$ released as visible/near-IR radiation.

The time dependence of the flash can be abstracted by an effective duration $\tau_{\rm eff}$ defined as
\begin{equation}
E_{\rm vis} = \int_{-\infty}^{+\infty} \phi(t)\,{\rm d}t, \qquad 
\tau_{\rm eff} \equiv \frac{E_{\rm vis}}{\phi_{\rm pk}},
\end{equation}
where $\phi(t)$ is the instantaneous radiant power and $\phi_{\rm pk}$ its peak.
If the lightcurve has total duration $T$ and a shape factor $\chi$ (e.g., $\chi=1$ for a rectangular pulse, $\chi=2$ for a triangular pulse), then
\begin{equation}
\tau_{\rm eff} = \frac{T}{\chi}, \qquad 
\phi_{\rm pk} = \chi\,\frac{E_{\rm vis}}{T}.
\end{equation}

The peak flux at Earth is then defined as $F_{\rm pk} = \frac{\phi_{\rm pk}}{4\pi \Delta^2}$, with $\Delta$ as the Earth–Moon distance. Restricting to the $V$ band, with fractional energy $f_V$ of the spectrum and zero-point flux $F_{V,0}$, the apparent peak magnitude is
\begin{equation}
m_V = -2.5\,\log_{10}\!\left(\frac{F_{V,{\rm pk}}}{F_{V,0}}\right), \qquad
F_{V,{\rm pk}} = f_V\,F_{\rm pk}.
\end{equation}

Applying this framework with $E_{\rm vis}\sim3\times10^{14}$ J and $T\sim100$–1000 s yields $\phi_{\rm pk}\sim10^{11}$–$10^{12}$ W, giving a peak flux at Earth of order $10^{-7}$–$10^{-6}$ W m$^{-2}$ and an apparent magnitude $m_V\simeq -2.5$--$-3$. To estimate the exact duration of the flash, we adopt the empirical equation \citep{madiedo2015analysis}
\begin{equation}
\tau = (77.6\pm34.4)\,\exp\!\bigl[-(0.94\pm0.06)\,m_V\bigr],
\end{equation}
which for $m_V\simeq -2.5$--$-3$ gives a duration of $\sim$300--2000 s, consistent with our initial assumption.

The spectral peak is near 1.2--1.6\,$\mu$m, corresponding to blackbody temperatures of 1800--2300 K. The spectral radiance of a blackbody \citep{bouley2012power} is
\begin{equation}
L(\lambda,T) = \frac{2\pi h c^{2}}{\lambda^{5}}
\left[\exp\!\left(\frac{hc}{\lambda k_{\rm B}T}\right)-1\right]^{-1},
\end{equation}
and the effective temperature $T_{\rm eff}$ is constrained by the energy balance
\begin{equation}
\frac{P_{0}}{\beta \pi R^{2}} = \int_{\lambda_{1}}^{\lambda_{2}} L(\lambda,T_{\rm eff})\,{\rm d}\lambda ,
\end{equation}
where $P_{0}$ is the peak radiant power, $\beta$ is a geometric correction factor, $R$ is the effective source radius, and $(\lambda_{1},\lambda_{2})$ denote the observational bandpass (here 0.4--0.9\,$\mu$m). Applying this relation yields $T_{\rm eff}\sim1800$--2300 K, consistent with the near-IR peak.

Following the plasma flash, shock heating produces a $\sim$100 m melt pool at $T\sim2000$ K. Radiative cooling calculations predict an infrared afterglow detectable for hours to days: effective temperatures decline from $\sim$1500 K to $\sim$500 K within several hours, but remain elevated ($\gtrsim300$ K) for 1 to 2 days. Such a km-scale anomaly would be accessible to infrared instruments on LRO/Diviner or large-aperture ground-based telescopes during lunar night.

\subsection{Seismic Shock and Moonquake}

A fraction of the impact energy transforms into seismic waves that cause shallow moonquakes with large magnitudes \citep{nunn2020lunar}. Following empirical scaling, the seismic energy is
\begin{equation}
E_{\rm seismic} = k\,E_{\rm imp},
\end{equation}
with efficiency $k\sim10^{-4}$. For 2024~YR\textsubscript{4} ($E_{\rm imp}\sim3\times10^{16}$\,J), this gives $E_{\rm seismic}\sim3\times10^{12}$\,J. The corresponding seismic magnitude can be estimated from the Gutenberg–Richter relation \citep{melosh1989impact}
\begin{equation}
M_{\rm seismic} = \frac{2}{3}\,\log_{10} E_{\rm seismic} - 3.2 ,
\end{equation}
which yields $M_{\rm seismic}\simeq5.0$--5.1. 

Impact‐induced seismic waves include compressional (P) and shear (S) body waves as well as surface Rayleigh waves. Because the Moon’s liquid core and fractured crust strongly attenuate S‐wave propagation, surface waves dominate the observable ground motion. For a given impactor impulse $mU$, the acceleration spectral density of the body waves can be estimated empirically \citep{nunn2024artificial} as
\begin{equation}
A_{\rm seismic} = 1.1\times10^{-12} S mU f^{3}/D ,
\end{equation}
where $A_{\rm seismic}$ is in m\,s$^{-2}$\,Hz$^{-1/2}$, $S$ is the seismic amplification (typically $S\!\sim\!1.7$), $mU$ is the impactor momentum (kg\,m\,s$^{-1}$), $f$ is the frequency (Hz), and $D$ is the epicentral distance (m). 

To estimate far‐field Rayleigh‐wave amplitudes from the impact source, we adopt the standard cylindrical geometric‐spreading and anelastic‐attenuation form \citep{aki2002quantitative,sato2012seismic}:
\begin{equation}
A(r,f) = A_{0}(f)\,r^{-1/2}\,\exp\!\left[-\frac{\pi f\,r}{Q_{R}\,v_{R}}\right],
\end{equation}
where $A_{0}(f)$ is the source amplitude at frequency $f$, $v_{R}$ is the Rayleigh‐wave speed ($\simeq2.6$ km s$^{-1}$), and $Q_{R}\!\sim\!10^{3}$ is the quality factor of the lunar crust \citep{blanchette2012investigation,garcia2011very}. The corresponding peak ground velocity (PGV) at a seismic station is
\begin{equation}
{\rm PGV}(r,f)= {\rm PGV}_{0}(f)\,r^{-1/2}\,
\exp\!\left[-\frac{\pi f\,r}{Q_{R}\,v_{R}}\right],
\end{equation}
where ${\rm PGV}_{0}(f)$ is the near‐source reference amplitude. These relations enable estimation of the signal strength at any lunar seismometer as a function of frequency and distance from the impact site.

The combined models indicate that even at farside distances ($r\!\gtrsim\!1500$ km), 2024~YR\textsubscript{4}‐type impacts would generate surface‐wave velocities above $10^{-6}$ m\,s$^{-1}$, readily detectable by current or planned lunar seismometers. Because of the Moon’s low attenuation and scattering, reverberations may persist for tens of minutes. Near the impact, ground motion would be extreme; globally, spectral amplitudes remain above Apollo noise levels, ensuring detection by any sensitive station. Secondary processes such as ejecta fallback and distal secondary impacts could also inject transient seismic energy. Overall, the predicted $M\!\sim\!5$ moonquake would rank among the strongest modern lunar seismic events, providing a rare opportunity to probe crustal structure, attenuation, and energy coupling in the lunar interior.

\begin{figure*}[ht!]
    \centering
    \includegraphics[width=0.7\linewidth]{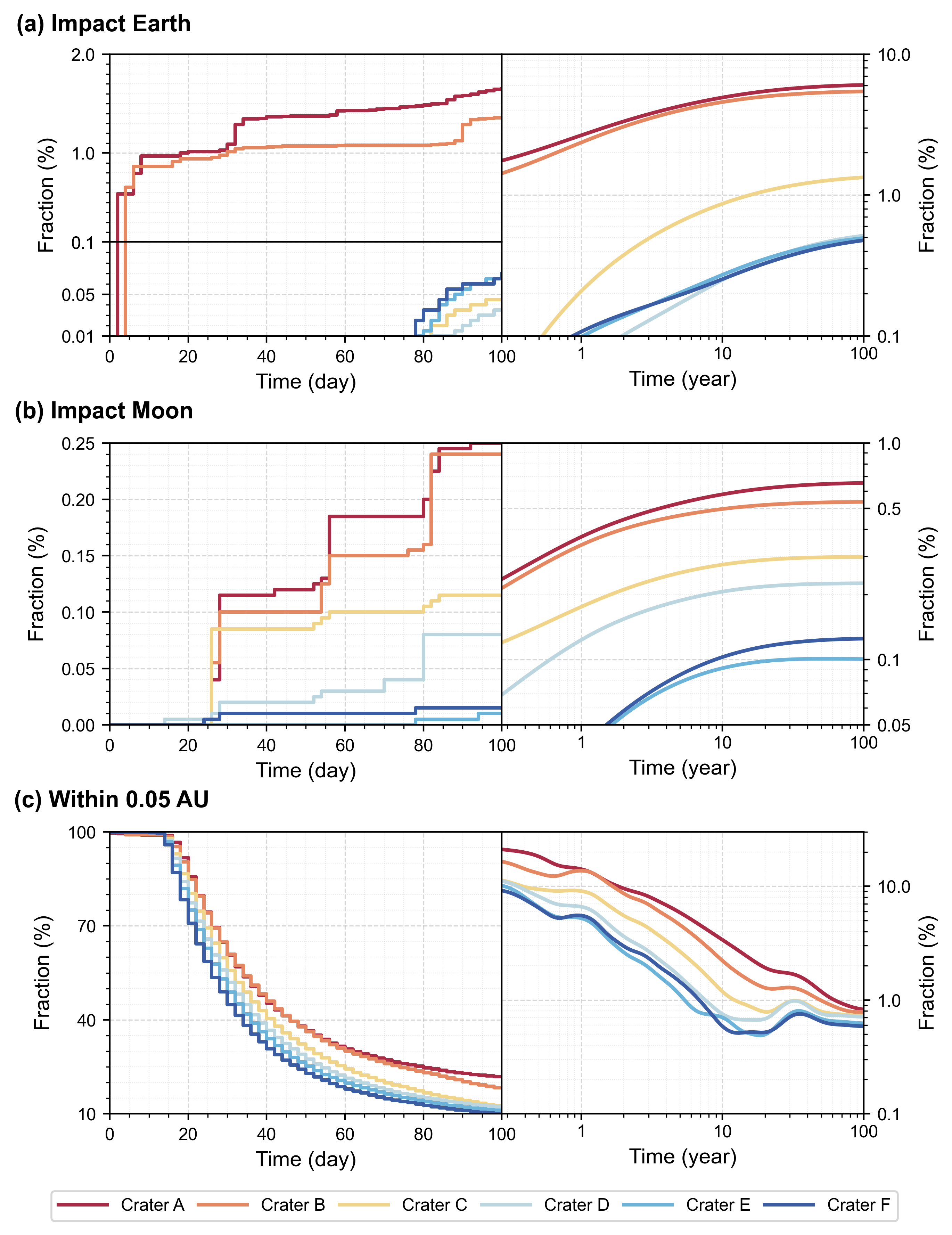} 
\caption{Time evolution of ejecta fractions impacting Earth (a) and the Moon (b), or remaining within 0.05 AU of Earth (c). The time axis is split into linear (first 100 days) and logarithmic (up to 100 yr) scales. In (a), the left panel uses a broken y-axis to better display the low fractions from Craters C–F. \label{fig:3}}
\end{figure*}

\subsection{Evolution of Escaped Lunar Ejecta}

The 2024~YR\textsubscript{4} impact is expected to eject $10^7\textendash10^8$ kg of lunar material that escapes the Moon (Section \ref{sec:3.1}), generating a stream of particles from millimeter-sized meteoroids to meter-scale boulders. Our orbital simulations reveal the diverse dynamical fates of these ejecta and highlight their strong dependence on the impact location on the Moon.

\subsubsection{Lunar ejecta flux to Earth}

For an impact on the lunar trailing side (i.e., Craters A and B), the dynamics favor a high Earth-delivery efficiency. 
Although the high impact angles ($60^\circ$--$84^\circ$) limit the total ejecta mass, the debris is launched with lower geocentric velocities, facilitating a rapid transfer. The first wave of particles arrives within 2--8~days, delivering $\sim$1\% of the total mass immediately. 
Based on the size distribution in Eq.~(\ref{eq:2}), the total ejecta mass ($4 \times 10^7$--$2 \times 10^8$~kg) yields up to $10^{13}$ meteoroids ($>$~1~mm), corresponding to an average naked-eye flux of $\sim$1--$6 \times 10^5$ hr$^{-1}$.
Furthermore, we estimate a fireball rate of 100--400 hr$^{-1}$, considering the luminosity expected at an entry speed near Earth's escape velocity as derived from our orbital simulations \citep{ceplecha1998meteor,subasinghe2018luminous}.
This intensity represents a historic storm posing severe hazards to satellites, a conclusion consistent with \citet{wiegert2025potential}. 
Following this initial burst, the flux peaks again at $\sim$30 and $\sim$90 days post-impact, with a cumulative 5--6\% of the escaped lunar debris entering Earth's atmosphere over 100 years.

In contrast, the latest orbit determination of 2024~YR\textsubscript{4} places the potential impact primarily on the Moon's leading side (i.e., Craters C--F). 
Ejecta from the leading side have higher geocentric velocities, causing a larger fraction to escape the Earth--Moon system (Figure \ref{fig:3}c) and reducing the Earth-delivery efficiency by an order of magnitude relative to the trailing side.
The earliest meteoroids reach Earth after $\sim$80~days, and Earth accretes only $\sim$1\% of the total mass over 100 years, which is comparable to the first-week delivery fraction from a trailing-side impact.
Notably, the leading-side impact cases of 2024~YR\textsubscript{4} are generally less oblique and could release more than $2 \times 10^8$~kg of ejecta, producing a meteor shower of up to $2 \times 10^7$ hr$^{-1}$ starting 80~days post-impact, despite the lower delivery ratio (Figure \ref{fig:3}a).

\subsubsection{Predicted Lunar Meteorite Delivery}

The ejecta flux includes 50--350 boulders of meter- sized (derived from Eq.~(\ref{eq:2}), corresponding to a total escaping mass of $10^5 \sim 10^6$ kg) that escape the Moon, depending on the impact location.
Upon entering Earth's atmosphere, these objects undergo catastrophic disruption, potentially yielding surviving meteorite masses of 0.1--3\% of the pre-atmospheric mass \citep{popova2011very}.

\begin{figure*}[ht!]
    \centering
    \includegraphics[width=0.9\linewidth]{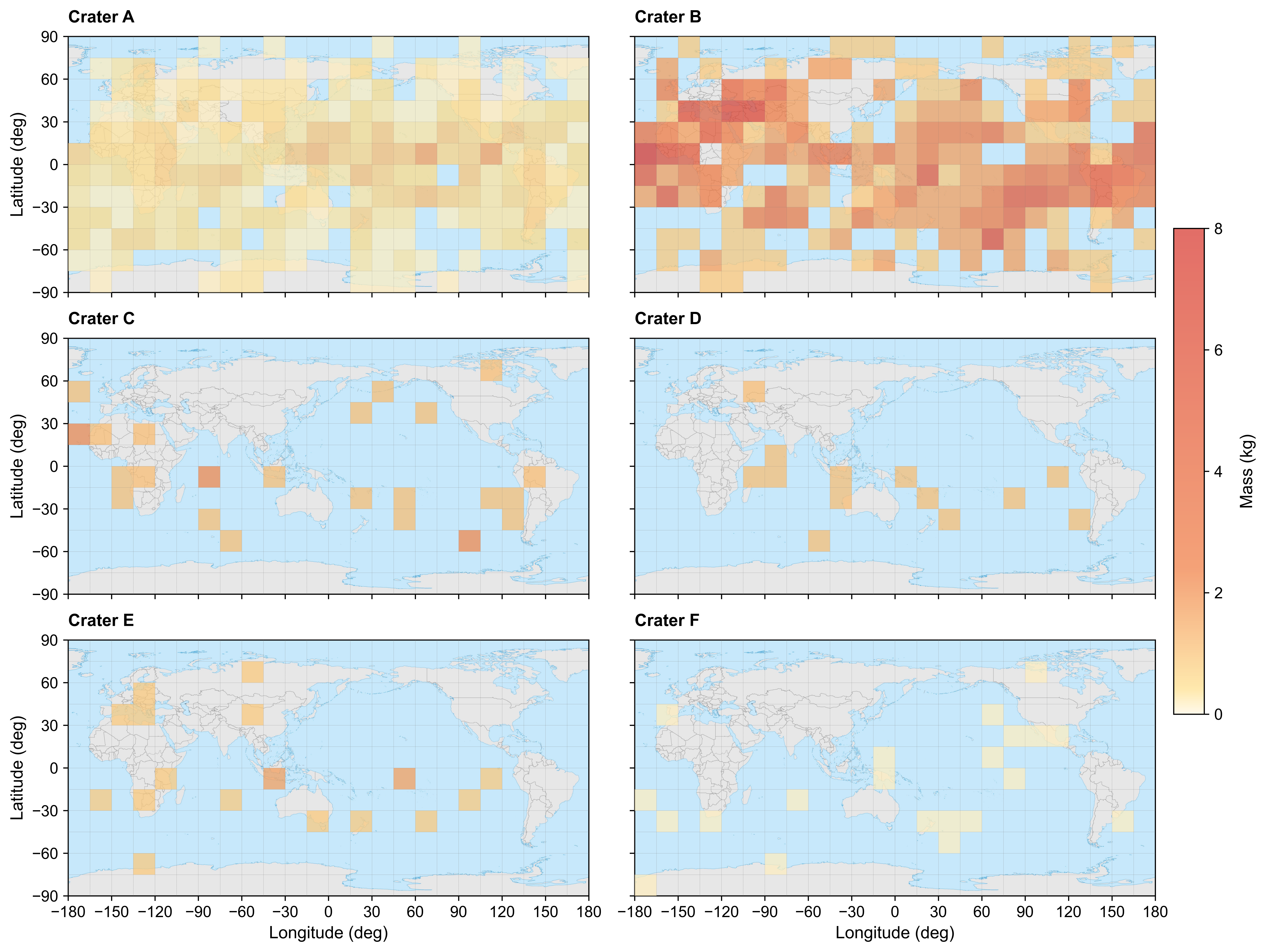} 
\caption{Predicted global distribution of surviving meteorite mass delivered to Earth over the first two years post-impact (T0 to T0 + 2~yr). The panels map the expected cumulative mass within each grid cell, projected onto a standard world map, for the six source craters.
\label{fig:4}}
\end{figure*}

By combining the initial boulder yield with the Earth-delivery probability derived from our orbital simulations, we map the expected meteorite fall flux within the first two years after the impact, projected onto a standard world map\footnote{\href{http://bzdt.ch.mnr.gov.cn/browse.html?picId=''4o28b0625501ad13015501ad2bfc0073''}{Standard Map Service, Map Approval No. GS(2016)1665.}} (Figure \ref{fig:4}).
The lunar trailing side (i.e., Craters A and B) offers the greatest potential for sample recovery due to its superior transport efficiency.
Specifically, an impact on Crater B maximizes the yield, delivering a total meteorite mass of $\sim$400~kg within the first year.
Conversely, impacts on the leading side (i.e., Craters C--F) result in a significantly lower yield, estimated at $\lesssim$20~kg over the same period, despite the potentially larger initial ejecta mass.

The simulations show that meteorite falls occur across all latitudes, confirming potential falls in Antarctica, which remains a prime location for meteorite recovery \citep{marvin1983discovery}. 
For the maximum-yield scenario (Crater B), the fallout concentrates along a diagonal band that extends across South America, North Africa, and the Arabian Peninsula. These regions contain extensive arid environments that are highly favorable for future search campaigns, offering a unique opportunity to recover samples with definite geological provenance.

\subsubsection{Secondary impacts on the Moon}

While some lunar ejecta strike Earth, others remaining bound to the Earth--Moon system may subsequently re-impact the Moon as they cross its orbital plane (Figure \ref{fig:3}b). 
Re-impact probabilities depend strongly on the source location, peaking at $\sim$0.25\% (within 100 days) for trailing-side ejecta while dropping significantly for the leading side.
The secondary impacts exhibit monthly peaks, a periodicity arising from fragments in geocentric orbits with near-lunar periods that repeatedly encounter the Moon at their orbital nodes.
Over 100 years, the cumulative re-impact mass is less than double the first-year value, indicating a strong concentration at early times.

With an assumed luminous efficiency of $2 \times 10^{-3}$, secondary impacts may produce observable optical flashes when their kinetic energy exceeds $10^6$ J, which corresponds to the $\sim$11.2 mag detection limit of NELIOTA \citep{xilouris2018neliota}.
Our orbital simulations, weighted by the ejecta size distribution, show that an impact on the trailing side (e.g., Crater B) produces $\gtrsim$2000 observable flashes within the first year. 
In contrast, impacts on the leading side show widely varying results: Craters C and D account for $\sim$800--1200 secondary flashes, while Craters E and F yield fewer than 200.
Beyond these orbital returns, the prompt fallback of massive, non-escaping fragments is expected to trigger numerous additional flashes immediately following the main impact.

\subsubsection{Detectability of meter-scale boulders} \label{sec:3.4.4}

Within the first 100 days, the fraction of ejecta remaining in Earth’s vicinity ($<$0.05 AU) drops to 10--20\%.
During the following century, this fraction gradually decreases and levels off at $\sim$1\%, reflecting a dynamic equilibrium between fragments drifting away and their returning (Figure \ref{fig:3}c).
This remaining fraction corresponds to a substantial population of meter-scale boulders near Earth, motivating us to assess their detectability with current surveys.

The Legacy Survey of Space and Time (LSST) can detect trailed sources down to magnitude $\sim$24 with angular velocities up to $\sim$10 deg day$^{-1}$ \citep{wu2025detectability,ivezic2019lsst}. 
We estimate the number of detectable 1--2~m boulders, assuming a minimum 10-day observational arc.
In high-yield scenarios (Craters B--E), LSST is expected to identify a total of 20--50 distinct objects over 100 years post-impact, whereas lower-yield cases (Craters A and F) produce $\lesssim$10. 
Among these detectable targets, the majority are observed while bound to Earth (distance $<0.01$ AU, geocentric energy $<0$).
This fraction exceeds 60\% in all cases (averaging $\sim$75\%) and reaches $>$80\% for the trailing-side craters. 
While most detections ($\sim$70\% averaged across all craters) occur within the first year, a distinct subset ($>$10\%) is identified upon returning from heliocentric orbits 50--100 years later, indicating that the 2024 YR\textsubscript{4} impact leaves an enduring observational signature in near-Earth space.
Crucially, our simulations indicate that $\sim70$\% of these targets overall remain observable for $>$0.1 years, allowing extended windows for precise orbit determination.

Extending the survey to the sub-meter regime (0.5--1~m) provides a valuable supplement by increasing the candidate count by a factor of $\sim$4--5. 
However, the fraction of detectable objects yielding arcs $>$0.1~years drops to $\sim$44\% at 0.5~m and merely $\sim$7\% at 0.3~m, below which detections become negligible.

\setlength{\tabcolsep}{1pt}
\newcommand{\sideshead}[1]{%
  \\[-4pt]
  \multicolumn{5}{l}{%
    \rule[0.5ex]{0.04\textwidth}{0.4pt}\hspace{0.5em}\textbf{#1}\hspace{0.5em}\rule[0.5ex]{0.04\textwidth}{0.4pt}
  }\\[1pt]
}

\begin{deluxetable*}{l c c c l l}
\tabletypesize{\scriptsize}
\tablewidth{0pt}
\tablecaption{Observability Timeline for the 2024 YR$_4$ Lunar Impact Event\label{tab:global_observability}}
\tablehead{
  \colhead{Observing Platform} &
  \colhead{Detection Window} &
  \colhead{Phenomenon} &
  \colhead{Peak Value} &
  \colhead{Notes} &
  \colhead{Science Qs.}
}
\startdata
\sideshead{Optical Flash Phase}
\shortstack[l]{Earth-based telescopes ($\ge$0.3\,m)} & $T_0$ to $+\sim10^3$\,s & Optical flash & $m_V\simeq-2.5$ to $-3$ &
\shortstack[l]{Visible from Hawaii} &
\shortstack[l]{Q1 (luminous efficiency, angle/velocity);\\ Q3 (early vapor lines)}
\\
Naked-eye observers & $T_0$ to $+\sim10$\,s & Optical flash & $m_V\simeq-3$ &
\shortstack[l]{Brief, star-like burst} &
\shortstack[l]{Q1 (occurrence, gross energy)}
\\
\shortstack[l]{JWST / space telescopes} & $T_0$ to $+\sim10^3$\,s & Optical/NIR flash & $m_{1.6\,\mu{\rm m}}\simeq-5$ &
\shortstack[l]{Coordinate ToO observation} &
\shortstack[l]{Q1 (radiative partition);\\ Q3 (vapor composition)}
\\
\shortstack[l]{Lunar orbiters (e.g., LRO)} & $T_0$ to +hours & Optical/IR plume & Bright transient &
\shortstack[l]{Direct line-of-sight required} &
\shortstack[l]{Q1 (plume kinematics);\\ Q3 (plume–exosphere coupling)}
\\[2pt]
\sideshead{Thermal Infrared Afterglow}
\shortstack[l]{Earth-based IR telescopes} & +10\,min to +48\,h & Thermal afterglow & 1500--500\,K &
\shortstack[l]{Favorable during lunar night} &
\shortstack[l]{Q3 (thermal inertia/porosity, melt volume);\\ Q1 (cooling energetics)}
\\
\shortstack[l]{Space IR (JWST/MIRI)} & +min to +days & Thermal emission & 1500--300\,K &
\shortstack[l]{Constrain melt composition} &
\shortstack[l]{Q3 (emissivity, volatiles)}
\\
\shortstack[l]{Lunar orbiters (Diviner, etc.)} & +min to +weeks & Surface heating & 1500--100\,K &
\shortstack[l]{High-res thermal crater mapping} &
\shortstack[l]{Q3 (T(x,y,t), layering, melt geometry);\\ Q1 (obliquity heating asymmetry)}
\\[2pt]
\sideshead{Seismic Shock Phase}
Lunar surface seismometers & $T_0$ to +hours & P, S, Rayleigh waves & $M_{\rm seismic}\simeq5.0$--5.1 &
\shortstack[l]{Global detection expected} &
\shortstack[l]{Q2 (source, $Q$, structure);\\ Q1 (coupling)}
\\
\shortstack[l]{\hspace{3mm}CLPS Farside Seismic Suite} & +8.0–12.0\,min, for 10–30 min & Rayleigh wave & PGV$\!\simeq$0.17\,mm\,s$^{-1}$ &
\shortstack[l]{VBB + SP array;\,easily detectable} &
\shortstack[l]{Q2 (farside attenuation/anisotropy)}
\\
\shortstack[l]{\hspace{3mm}Artemis LEMS (South Pole)} & +11–12\,min, for 10–30 min & Rayleigh wave & PGV$\!\simeq$0.09\,mm\,s$^{-1}$ &
\shortstack[l]{High S/N;\,autonomous polar station} &
\shortstack[l]{Q2 (polar structure, dispersion)}
\\
\shortstack[l]{\hspace{3mm}Chang’e 8 lander} & +10–13\,min, for 10–30 min & Rayleigh wave & PGV$\!\simeq$0.10\,mm\,s$^{-1}$ &
\shortstack[l]{Clear signal;\,seismometer onboard} &
\shortstack[l]{Q2 (near-side megaregolith)}
\\
\shortstack[l]{\hspace{3mm}LUPEX (JAXA–ISRO)} & +10–13\,min, for 10–30 min & Rayleigh wave & PGV$\!\simeq$0.10\,mm\,s$^{-1}$ &
\shortstack[l]{Detectable if seismic package onboard} &
\shortstack[l]{Q2 (regional $Q$, dispersion)}
\\
\shortstack[l]{\hspace{3mm}South-Pole Station (85\degr S, 0\degr E)} & +8.9\,min, for 10--30 min & Rayleigh wave & PGV$\!\simeq$0.16\,mm\,s$^{-1}$ &
\shortstack[l]{Broadband; favorable for polar missions} &
\shortstack[l]{Q2 (polar tomography)}
\\
\shortstack[l]{Lunar orbiters with seismic payload} & +min to +hours & Orbital acceleration & Moderate &
\shortstack[l]{Possible small orbital perturbation} &
\shortstack[l]{Q2 (long-period response)}
\\[2pt]
\sideshead{Escaped Lunar Ejecta}
\shortstack[l]{Ground meteor networks} & +2~d to 100~d & Meteor storm & $\sim 5 \times 10^8$ h$^{-1}$ &
\shortstack[l]{Severe satellite hazard} &
\shortstack[l]{Q1 (escape fraction, size–speed);\\ Q3 (cislunar dust environment)}
\\
\shortstack[l]{Meteorite search campaigns} & +weeks to 1 yr & Meteorite falls & $\sim 400$ kg (Total) &
\shortstack[l]{Antarctica/Arid regions} &
\shortstack[l]{Q3 (shock/volatiles/composition link);\\ Q1 (ejecta provenance)}
\\
\shortstack[l]{Flash monitoring (NELIOTA)} & +days to months & Secondary impacts & $\sim 3000$ flashes &
\shortstack[l]{} &
\shortstack[l]{Q1 (recapture efficiency);\\ Q3 (gardening rate)}
\\
\shortstack[l]{Optical wide-field survey (LSST)} & 0.1 to 100 yr & Boulders ($\sim1$ m) & 20--50 objects &
\shortstack[l]{Accessible mission targets} &
\shortstack[l]{Q1 (three-body pathways, RP response);\\ Q3 (fresh boulder spectra)}
\\
\enddata
\tablecomments{T$_0$ = nominal impact epoch (2032 Dec 22 15:19 UTC).
Optical/IR parameters assume $\eta=10^{-2}$; seismic amplitudes from $E_{\rm seismic}=kE_{\rm imp}$ with $k=10^{-4}$ and
Rayleigh-wave attenuation ${\rm PGV}(r,f)={\rm PGV}_0(f)\,r^{-1/2}\exp[-\pi f r/(Q_R v_R)]$ using $v_R=2.6$ km s$^{-1}$, $Q_R=1000$.
Distances $r$ are great-circle separations from the impact site (35.5\degr E, 40\degr S).
Peak values in section ``Escaped Lunar Ejecta'' correspond to the maximum-yield scenario (Crater B on the trailing side).
\textbf{Science question tags:} \textbf{Q1} = Impact dynamics (energy partition, incidence/velocity, ejecta angles–speeds, escape/recapture, three-body evolution). \textbf{Q2} = Lunar interior \& response (source time function, attenuation/scattering $Q$, dispersion, crust/megaregolith structure, normal modes). \textbf{Q3} = Surface/subsurface materials (thermophysical properties, melt/glass/breccia, volatile retention, boulder SFD, composition links incl.\ meteorites).
}
\end{deluxetable*}

\section{Observation Timeline}\label{sec:highlight}

Given the diversity of physical phenomena predicted—from optical flash to infrared afterglow, seismic shock, and delayed meteoroid flux—a coordinated, multi-platform campaign is essential to capture the 2024~YR\textsubscript{4} lunar impact comprehensively. We have been devoted to apply a systematic analytical framework to the specific 2024~YR\textsubscript{4} lunar impact to probe the Moon as \citet{jiao2025probing}. Table~\ref{tab:global_observability} summarizes all major observables and their detection windows across optical, infrared, seismic, and dynamical domains. Below, we outline an integrated observing strategy following the natural temporal sequence of the event.

In brief, each observable targets distinct science: the optical flash (rise/decay and spectra) constrains kinetic energy, impact angle/velocity, luminous efficiency, and early vapor composition; the NIR/MIR afterglow (cooling curves and maps) probes melt volume, glass/breccia fractions, and regolith thermophysical properties (thermal inertia, porosity, blockiness) under known boundary conditions; broadband seismic wavefields (Rayleigh, body waves, normal modes) recover the source time function and test crust–megaregolith structure, attenuation, scattering, and near-surface layering; orbital imaging and thermal mapping (LROC/Diviner) close the loop on cratering scalings in the km regime, ejecta asymmetry versus obliquity/azimuth, boulder size–frequency distributions, and melt morphology; Earth-adjacent meteoroid flux (mm–cm) diagnoses the high-velocity tail, escape fraction, and the transient cislunar dust environment—also calibrating faint-impact detectability for planetary-defense sensors; meter-class ejecta tracking tests three-body dynamics and radiation-pressure response of fresh lunar boulders while enabling spectroscopy of “lunar-origin mini-asteroids”; secondary lunar impacts measure ejecta recapture efficiency and regolith gardening rates; and recovered meteorites (Fig.~\ref{fig:4}) provide ground-truth on shock metamorphism, volatile retention, and isotopic/compositional links between the cratered surface, the ejecta field, and small-body provenance.

\subsection{Pre-Impact Preparations}

Well before the final approach, additional astrometric opportunities will arise that can substantially refine the orbit of 2024~YR\textsubscript{4}. In particular, a favorable Earth encounter and observational window around 2028 is expected to enable high-precision optical and radar tracking, significantly reducing uncertainties in the impact probability and corridor on the Moon. These pre-2032 observations will be decisive in confirming whether a lunar impact will occur and in narrowing the predicted impact geometry to a level suitable for targeted observation planning.

In the months leading to the event, precise refinement of 2024~YR\textsubscript{4}’s orbit is critical to constrain the impact corridor and timing to within minutes. Ephemeris refinement will determine which lunar hemisphere and longitude sector to monitor \citep{suggs2008nasa, xilouris2018neliota, bonanos2015neliota}. Large telescopes (8–10\,m class and beyond) can allocate short-term slots around the expected epoch, while networks of medium telescopes (0.3–1\,m) with high-speed photometers should plan continuous video coverage to ensure redundancy \citep{slade2010goldstone, ding2024china}. On the Moon, active orbiters and landers should be placed in recording or imaging mode, and seismometers armed for immediate data logging. Space agencies can coordinate orbital adjustments of LRO, Chandrayaan-3, or other platforms to enable rapid post-impact passes \citep{savage1996interstellar, sabelhaus2004overview, gong2019cosmology, chowdhury2020imaging}.  

The predicted impact on 2032-12-22 occurs when the Moon is waning gibbous ($\sim$70\% illumination). Geometry analysis shows that the southern impact corridor will be visible across the Pacific hemisphere of Earth \citep{merisio2023present}. In particular, the Mauna~Kea observatories (Keck, Subaru, Gemini-N) will enjoy optimal visibility with the Moon high ($\sim$80°) above the horizon and local pre-dawn darkness. Western North America will see the Moon near dawn at lower elevation, while South America, Europe, and Asia will have daylight or low Moon, rendering the event unobservable from those regions. Thus, coordinated efforts from Hawaii-based facilities, supported by global amateur networks, offer the best chance for optical confirmation.

\subsection{Moment of Impact (T\textsubscript{0})}

The first seconds are dominated by the optical flash. Observers should begin continuous imaging several minutes prior to the nominal time to capture the rise phase. The expected brightness of $m_V\sim$–3 to –5 makes the event potentially visible to the naked eye if occurring on the lunar night side (with a extremely low probability) \citep{madiedo2019multiwavelength}, though contrast against a sunlit surface may reduce detectability but remain strong visibility to hypersensitized telescopes \citep{yanagisawa2006first, madiedo2018first}. High-speed photometry (10–100\,Hz cadence) is essential to resolve the sub-second rise and multi-second decay of the flash. If available, large-aperture telescopes can attempt rapid spectroscopy to identify emission lines of vaporized silicates or metals. In parallel, orbital imagers (e.g., LROC or small CubeSat payloads) can monitor the near field, while near-infrared facilities could detect the same flash if rapid target-of-opportunity (ToO) observations are enabled.  

Even if the impact site lies just beyond the terminator, the event may still illuminate the surrounding terrain or generate Earthshine variations detectable from ground-based photometry.  

\subsection{Minutes to Hours after T\textsubscript{0}}

Once the flash fades, the thermal phase begins. Infrared observations should commence within minutes to record the cooling curve of the newly formed $\sim$1\,km crater. Ground-based mid-IR telescopes (e.g., IRTF, Gemini-North) \citep{deutsch2003mirsi} can track the afterglow from $\sim$1500\,K to a few hundred Kelvins over several hours, while space observatories such as JWST/MIRI can extend this monitoring to longer wavelengths and finer sensitivity \citep{rieke2015mid}. LRO’s Diviner instrument provides the ideal asset for mapping the temperature evolution and identifying melt deposits \citep{williams2017global, paige2010diviner}; follow-up imaging by LROC in the ensuing daylight will reveal the fresh crater morphology and ejecta rays \citep{robinson2010lunar}.  

Simultaneously, lunar seismometers will register the shock. Rayleigh waves are expected to reach near-side Apollo sites within 7–15\,minutes, with predicted peak ground velocities of $10^{-2}$–$10^{-1}$\,mm\,s$^{-1}$—well above historic Apollo noise levels \citep{garcia2011very, blanchette2012investigation}. Even farside or polar stations should detect clear arrivals, offering a rare opportunity to probe the Moon’s crustal structure using a known energy source. 

Finally, lunar seismic records from all stations will be analyzed to reconstruct source parameters, verify energy coupling efficiency, and refine models of lunar interior structure. The 2024~YR\textsubscript{4} impact, if it occurs, would thus provide a complete multi-modal dataset: optical and infrared emissions tracing energy release, seismic waves probing the interior, and orbital imaging revealing geological consequences—a natural experiment in planetary impact physics.

\subsection{Days to Years after T\textsubscript{0}}

In the following days, attention shifts from immediate transients to slower processes and byproducts. As the impact site rotates into sunlight, lunar orbiters should obtain high-resolution optical imaging to measure the final crater dimensions and ejecta albedo. On Earth, meteor and radar networks should monitor for a transient increase in meteoroid flux roughly 2–8\,days post-impact, corresponding to lunar ejecta fragments reaching Earth’s vicinity. These fragments—typically mm–cm in size—could produce a short-lived meteor shower radiating from the Moon’s orbital direction. 


The first month after $T_0$ is the high-yield window for targeted campaigns. Accordingly, ground-based meteor cameras and high-power radars should maintain near-continuous coverage from $T_0{+}2$ to $T_0{+}10$\,days to capture the early peak in the trailing-side cases, then intensify cadence around $T_0{+}30$ and $T_0{+}90$\,days for secondary peaks. For leading-side impacts, the same infrastructure should pivot to a deferred campaign starting near $T_0{+}80$\,days and extending for several weeks, with expectations set for lower but detectable flux.

We recommend a tiered follow-up: wide-field discovery and alert brokering; rapid astrometric recovery with 2–4\,m telescopes to refine orbits; prompt low-resolution reflectance spectroscopy on 4–10\,m class telescopes to test lunar provenance; and, where feasible, deep photometry or thermal IR (including space assets) to constrain size, albedo, and surface state. More specific observing classification has been listed in \citet{devogele2025rapid}.

A small but predictable fraction of escapees should re-impact the Moon with inter-peak spacings near the synodic period, favoring early-time returns (first few tens of days), which implies a relatively well-defined observation window for more concentrated observing resources. Although individual flashes are faint, a few-day continuous monitoring campaign around predicted windows—using existing lunar flash detection networks and small telescopes pointed to the appropriate selenographic longitudes—can directly test ejecta recapture physics and constrain the near-Moon particulate environment.

Model-guided recovery of lunar meteorites is also warranted. Orbit integrations coupled to size–velocity distributions predict that a small but non-negligible fraction of escaping fragments survive atmospheric entry as meteorites within the first two years. For trailing-hemisphere (east-longitude) impacts, the bulk of the delivery occurs within the first year with an order-of-magnitude cumulative mass in the $10^1$–$10^2$\,kg range and an upper envelope reaching a few $\times 10^2$\,kg under high-yield realizations; for leading-hemisphere (west-longitude) impacts, arrivals are delayed (commencing near $T_0{+}80$\,days) and the integrated mass is lower by a factor of a few. Figure~\ref{fig:4} summarizes these outcomes by showing the predicted global distribution of surviving meteorite mass delivered to Earth over the first two years following the impact ($T_0$ to $T_0{+}2$\,yr). For each of the six representative source craters, the panels map the cumulative recovered mass (kg) within equal-area geographic grid cells projected onto a standard world map. In all cases, the surviving mass is preferentially delivered to low- and mid-latitudes (typically within $\pm30^\circ$), with broad longitudinal coverage reflecting Earth’s rotation during debris interception. This geographic concentration implies that coordinated, regionally targeted recovery strategies—combining global meteor networks for trajectory reconstruction with rapid-response ground searches—can substantially enhance the probability of meteorite recovery. Fresh lunar meteorites from this event would anchor laboratory studies of regolith thermophysics and shock metamorphism and provide a compositional cross-check against remotely sensed spectra, directly linking the surface, the ejecta dynamics, and samples in hand.

As indicated in Section \ref{sec:3.4.4}, the year following the 2024 YR\textsubscript{4} impact offers a unique window to monitor tens to hundreds of observable meter- and sub-meter-sized lunar ejecta. 
While significantly smaller than typical observable Near-Earth Objects (NEOs), these boulders possess substantial mass.
Crucially, the majority remain bound to Earth, implying close proximity and low relative velocities. 
This accessibility presents two time-sensitive opportunities that warrant immediate mission planning: 
\begin{enumerate}
    \item Sample Retrieval: Capturing and maneuvering these fragments into stable geocentric orbits is feasible \citep{baoyin2010capturing}, allowing for the retrieval of massive samples originating from the lunar interior.
    \item Planetary Defense: These optimally sized targets serve as ideal testbeds for low-cost validation of defense technologies, such as conducting realistic kinetic impact experiments \citep{jiao2023optimal,lee2025investigation,lee2025trajectories}.
\end{enumerate}

\section{CONCLUSION}

If the potential lunar impact of asteroid 2024~YR\textsubscript{4} indeed occurs, it would produce an exceptionally bright, multi-modal signal and a once-in-$10^4$-year natural experiment on the Moon. According to our modeling, we may then expect an optical/NIR flash reaching $m_V\!\sim\!-3$ over $\sim10^2$–$10^3$\,s with $T_{\rm eff}\!\sim\!1.8$–$2.3$$\times 10^3$\,K, a global $M\!\sim\!5$ moonquake, a km-scale fresh crater, and a total mass of $10^{7}$–$10^{8}$\,kg of high-velocity ejecta linking the lunar surface to cislunar space.
These predictions finally yield a single, time-ordered plan from the first seconds to the following weeks.%

Beyond establishing the detectability, this event unlocks a broad science return that is rarely available in planetary science: 
(1) the surface and near-surface thermophysics under known boundary conditions (flash spectra, melt-pool cooling, Diviner-class thermal mapping); 
(2) the interior structure and attenuation from a calibrated, high-S/N Rayleigh-wave source recorded by modern stations (e.g., polar and farside packages), enabling joint inversions with legacy Apollo constraints; 
(3) the hypervelocity cratering, the energy coupling, and the oblique-impact ejecta systematics at kilometer scale for testing scaling laws and numerical models; 
and (4) source-tagged dust and meteoroids that trace regolith–exosphere exchange and deliver lunar material to near-Earth space—linking small-body provenance, lunar meteorites, and operational space-safety questions.%

Practically, our results provide a clear picture of the cross-platform campaign: high-speed photometry and rapid spectroscopy at $T_0$; hours-to-days mid-IR monitoring of the afterglow; immediate, networked seismic acquisition for global arrivals; rapid-response orbital imaging and thermal mapping; and short-lag radar/meteor surveillance for escaped ejecta. 
If the collision indeed happens, the 2024~YR\textsubscript{4} impact will become the brightest and best-characterized lunar impact on record and also a benchmark for impact and hazard models. 
If it does not, the methodology, visibility analysis, and coordination logic presented here remain a reusable blueprint for rapid-response observations of future natural impacts—turning potential hazard into preparedness and, ultimately, into discovery.


\begin{acknowledgments}

This work is supported by the National Natural Science Foundation of China under Grant Nos. 12572404, 62227901, U24B2048, 125B1015, 123B2038, the national level fund No. KJSP2023020301, the Beijing Nova Program No. 20250484831, and the Tsinghua Dushi Funds.

\end{acknowledgments}

\bibliography{sample701}{}

@article{jiao2024asteroid,
  title={Asteroid Kamo ‘oalewa’s journey from the lunar Giordano Bruno crater to Earth 1: 1 resonance},
  author={Jiao, Yifei and Cheng, Bin and Huang, Yukun and Asphaug, Erik and Gladman, Brett and Malhotra, Renu and Michel, Patrick and Yu, Yang and Baoyin, Hexi},
  journal={Nature Astronomy},
  volume={8},
  number={7},
  pages={819--826},
  year={2024},
  publisher={Nature Publishing Group UK London},
  doi={10.1038/s41550-024-02258-z}
}

@article{jiao2024sph,
  title={SPH--DEM modelling of hypervelocity impacts on rubble-pile asteroids},
  author={Jiao, Yifei and Yan, Xiaoran and Cheng, Bin and Baoyin, Hexi},
  journal={Monthly Notices of the Royal Astronomical Society},
  volume={527},
  number={4},
  pages={10348--10357},
  year={2024},
  publisher={Oxford University Press},
  doi={10.1093/mnras/stad3888}
}

@article{rein2012rebound,
  title={REBOUND: an open-source multi-purpose N-body code for collisional dynamics},
  author={Rein, Hanno and Liu, S-F},
  journal={Astronomy \& Astrophysics},
  volume={537},
  pages={A128},
  year={2012},
  publisher={EDP Sciences},
  doi={10.1051/0004-6361/201118085}
}

@article{burns1979radiation,
  title={Radiation forces on small particles in the solar system},
  author={Burns, Joseph A and Lamy, Philippe L and Soter, Steven},
  journal={Icarus},
  volume={40},
  number={1},
  pages={1--48},
  year={1979},
  publisher={Elsevier},
  doi={10.1016/0019-1035(79)90050-2}
}

@article{jiao2023optimal,
  title={Optimal kinetic-impact geometry for asteroid deflection exploiting delta-v hodograph},
  author={Jiao, Yifei and Cheng, Bin and Baoyin, Hexi},
  journal={Journal of Guidance, Control, and Dynamics},
  volume={46},
  number={2},
  pages={382--389},
  year={2023},
  publisher={American Institute of Aeronautics and Astronautics},
  doi={10.2514/1.G006876}
}

@article{ivezic2019lsst,
  title={LSST: from science drivers to reference design and anticipated data products},
  author={Ivezi{\'c}, {\v{Z}}eljko and Kahn, Steven M and Tyson, J Anthony and Abel, Bob and Acosta, Emily and Allsman, Robyn and Alonso, David and AlSayyad, Yusra and Anderson, Scott F and Andrew, John and others},
  journal={The Astrophysical Journal},
  volume={873},
  number={2},
  pages={111},
  year={2019},
  publisher={IOP Publishing},
  doi={10.3847/1538-4357/ab042c}
}

@article{lee2025investigation,
  title={Investigation of the incremental benefits of eccentric collisions in kinetic deflection of potentially hazardous asteroids},
  author={Lee, Kinthong and Fang, Zhengqing and Wang, Zhaokui},
  journal={Icarus},
  volume={425},
  pages={116312},
  year={2025},
  publisher={Elsevier},
  doi={10.1016/j.icarus.2024.116312}
}

@ARTICLE{wiegert2025potential,
       author = {{Wiegert}, Paul and {Brown}, Peter and {Lopes}, Jack and {Connors}, Martin},
        title = "{The Potential Danger to Satellites due to Ejecta from a 2032 Lunar Impact by Asteroid 2024 YR$_{4}$}",
      journal = {\apjl},
         year = 2025,
        month = sep,
       volume = {990},
       number = {1},
          eid = {L20},
        pages = {L20},
          doi = {10.3847/2041-8213/adfa8b},
archivePrefix = {arXiv},
       eprint = {2506.11217},
 primaryClass = {astro-ph.EP},
       adsurl = {https://ui.adsabs.harvard.edu/abs/2025ApJ...990L..20W}
}

@article{jiao2025probing,
  title={Probing the Moon from Future Asteroid Impacts: A Review},
  author={Jiao, Yifei and Cheng, Bin and Baoyin, Hexi},
  journal={arXiv preprint arXiv:2509.01436},
  year={2025}
}

@article{rivkin2025jwst,
  title={JWST Observations of Potentially Hazardous Asteroid 2024 YR4},
  author={Rivkin, AS and Mueller, T and MacLennan, E and Holler, B and Burdanov, A and de Wit, J and Pravec, P and Micheli, M and Devogele, M and Conversi, L and others},
  journal={Research Notes of the AAS},
  volume={9},
  number={4},
  pages={70},
  year={2025},
  publisher={The American Astronomical Society}
}

@article{madiedo2014large,
  title={A large lunar impact blast on 2013 September 11},
  author={Madiedo, Jos{\'e} M and Ortiz, Jos{\'e} L and Morales, Nicol{\'a}s and Cabrera-Ca{\~n}o, Jes{\'u}s},
  journal={Monthly Notices of the Royal Astronomical Society},
  volume={439},
  number={3},
  pages={2364--2369},
  year={2014},
  publisher={Oxford University Press}
}

@article{gudkova2011large,
  title={Large impacts detected by the Apollo seismometers: Impactor mass and source cutoff frequency estimations},
  author={Gudkova, TV and Lognonn{\'e}, PH and Gagnepain-Beyneix, J},
  journal={Icarus},
  volume={211},
  number={2},
  pages={1049--1065},
  year={2011},
  publisher={Elsevier}
}

@article{liu2025collision,
  title={Collision probability analysis of 2024 YR4},
  author={Liu, Xin and Hou, Xiyun and Cheng, Haowen},
  journal={npj Space Exploration},
  volume={1},
  number={1},
  pages={4},
  year={2025},
  publisher={Nature Publishing Group UK London}
}

@article{collins2004modeling,
  title={Modeling damage and deformation in impact simulations},
  author={Collins, Gareth S and Melosh, H Jay and Ivanov, Boris A},
  journal={Meteoritics \& Planetary Science},
  volume={39},
  number={2},
  pages={217--231},
  year={2004},
  publisher={Wiley Online Library}
}

@article{jutzi2015sph,
  title={SPH calculations of asteroid disruptions: the role of pressure dependent failure models},
  author={Jutzi, Martin},
  journal={Planetary and space science},
  volume={107},
  pages={3--9},
  year={2015},
  publisher={Elsevier}
}

@article{melosh1989impact,
  title={Impact Cratering: A Geologic Process Oxford Univ},
  author={Melosh, HJ},
  journal={Press, New York},
  year={1989}
}

@article{jutzi2008numerical,
  title={Numerical simulations of impacts involving porous bodies: I. Implementing sub-resolution porosity in a 3D SPH hydrocode},
  author={Jutzi, Martin and Benz, Willy and Michel, Patrick},
  journal={Icarus},
  volume={198},
  number={1},
  pages={242--255},
  year={2008},
  publisher={Elsevier}
}

@article{jutzi2009numerical,
  title={Numerical simulations of impacts involving porous bodies: II. Comparison with laboratory experiments},
  author={Jutzi, Martin and Michel, Patrick and Hiraoka, Kensuke and Nakamura, Akiko M and Benz, Willy},
  journal={Icarus},
  volume={201},
  number={2},
  pages={802--813},
  year={2009},
  publisher={Elsevier}
}

@article{collins2011size,
  title={The size-frequency distribution of elliptical impact craters},
  author={Collins, GS and Elbeshausen, D and Davison, TM and Robbins, SJ and Hynek, BM},
  journal={Earth and Planetary Science Letters},
  volume={310},
  number={1-2},
  pages={1--8},
  year={2011},
  publisher={Elsevier}
}

@article{luo2022ejecta,
  title={Ejecta pattern of oblique impacts on the Moon from numerical simulations},
  author={Luo, Xi-Zi and Zhu, Meng-Hua and Ding, Min},
  journal={Journal of Geophysical Research: Planets},
  volume={127},
  number={11},
  pages={e2022JE007333},
  year={2022},
  publisher={Wiley Online Library}
}

@article{merisio2023present,
  title={Present-day model of lunar meteoroids and their impact flashes for LUMIO mission},
  author={Merisio, Gianmario and Topputo, Francesco},
  journal={Icarus},
  volume={389},
  pages={115180},
  year={2023},
  publisher={Elsevier}
}

@article{madiedo2015analysis,
  title={Analysis of Moon impact flashes detected during the 2012 and 2013 Perseids},
  author={Madiedo, Jos{\'e} M and Ortiz, Jos{\'e} L and Organero, Faustino and Ana-Hern{\'a}ndez, Leonor and Fonseca, Fernando and Morales, Nicol{\'a}s and Cabrera-Ca{\~n}o, Jes{\'u}s},
  journal={Astronomy \& Astrophysics},
  volume={577},
  pages={A118},
  year={2015},
  publisher={EDP Sciences}
}

@article{ait2015first,
  title={First lunar flashes observed from Morocco (ILIAD Network): Implications for lunar seismology},
  author={Ait Moulay Larbi, Mamoun and Daassou, Ahmed and Baratoux, David and Bouley, Sylvain and Benkhaldoun, Zouhair and Lazrek, Mohamed and Garcia, Raphael and Colas, Francois},
  journal={Earth, Moon, and Planets},
  volume={115},
  number={1},
  pages={1--21},
  year={2015},
  publisher={Springer}
}

@article{bouley2012power,
  title={Power and duration of impact flashes on the Moon: Implication for the cause of radiation},
  author={Bouley, Sylvain and Baratoux, David and Vaubaillon, Jeremie and Mocquet, A and Le Feuvre, Mathieu and Colas, Francois and Benkhaldoun, Zouhair and Daassou, Ahmed and Sabil, Mohammed and Lognonn{\'e}, Philippe},
  journal={Icarus},
  volume={218},
  number={1},
  pages={115--124},
  year={2012},
  publisher={Elsevier}
}

@article{nunn2020lunar,
  title={Lunar seismology: A data and instrumentation review},
  author={Nunn, Ceri and Garcia, Raphael F and Nakamura, Yosio and Marusiak, Angela G and Kawamura, Taichi and Sun, Daoyuan and Margerin, Ludovic and Weber, Renee and Drilleau, M{\'e}lanie and Wieczorek, Mark A and others},
  journal={Space Science Reviews},
  volume={216},
  number={5},
  pages={89},
  year={2020},
  publisher={Springer}
}

@article{nunn2024artificial,
  title={Artificial impacts on the moon: Modeling 3D seismic propagation effects with AxiSEM3D},
  author={Nunn, Ceri and Fernando, Benjamin A and Panning, Mark P},
  journal={The Planetary Science Journal},
  volume={5},
  number={11},
  pages={246},
  year={2024},
  publisher={IOP Publishing}
}

@book{aki2002quantitative,
  title={Quantitative seismology},
  author={Aki, Keiiti and Richards, Paul G},
  year={2002}
}

@book{sato2012seismic,
  title={Seismic wave propagation and scattering in the heterogeneous earth},
  author={Sato, Haruo and Fehler, Michael C and Maeda, Takuto},
  volume={496},
  year={2012},
  publisher={Springer}
}

@article{blanchette2012investigation,
  title={Investigation of scattering in lunar seismic coda},
  author={Blanchette-Guertin, J-F and Johnson, Catherine L and Lawrence, Jesse F},
  journal={Journal of Geophysical Research: Planets},
  volume={117},
  number={E6},
  year={2012},
  publisher={Wiley Online Library}
}

@article{garcia2011very,
  title={Very preliminary reference Moon model},
  author={Garcia, Rapha{\"e}l F and Gagnepain-Beyneix, Jeannine and Chevrot, S{\'e}bastien and Lognonn{\'e}, Philippe},
  journal={Physics of the Earth and Planetary Interiors},
  volume={188},
  number={1-2},
  pages={96--113},
  year={2011},
  publisher={Elsevier}
}

@article{suggs2008nasa,
  title={The NASA lunar impact monitoring program},
  author={Suggs, Robert M and Cooke, William J and Suggs, Ronnie J and Swift, Wesley R and Hollon, Nicholas},
  journal={Earth, Moon, and Planets},
  volume={102},
  number={1},
  pages={293--298},
  year={2008},
  publisher={Springer}
}

@article{xilouris2018neliota,
  title={NELIOTA: The wide-field, high-cadence, lunar monitoring system at the prime focus of the Kryoneri telescope},
  author={Xilouris, EM and Bonanos, AZ and Bellas-Velidis, I and Boumis, P and Dapergolas, A and Maroussis, A and Liakos, A and Alikakos, I and Charmandaris, V and Dimou, G and others},
  journal={Astronomy \& Astrophysics},
  volume={619},
  pages={A141},
  year={2018},
  publisher={EDP Sciences}
}

@article{bonanos2015neliota,
  title={NELIOTA: ESA's new NEO lunar impact monitoring project with the 1.2 m telescope at the National Observatory of Athens},
  author={Bonanos, AZ and Xilouris, Manolis and Boumis, Panos and Bellas-Velidis, Ioannis and Maroussis, Athanasios and Dapergolas, Anastasios and Fytsilis, Anastasios and Charmandaris, Vassilis and Tsiganis, Kleomenis and Tsinganos, Kanaris},
  journal={Proceedings of the International Astronomical Union},
  volume={10},
  number={S318},
  pages={327--329},
  year={2015},
  publisher={Cambridge University Press}
}

@article{madiedo2019multiwavelength,
  title={Multiwavelength observations of a bright impact flash during the 2019 January total lunar eclipse},
  author={Madiedo, Jos{\'e} M and Ortiz, Jos{\'e} L and Morales, Nicol{\'a}s and Santos-Sanz, Pablo},
  journal={Monthly Notices of the Royal Astronomical Society},
  volume={486},
  number={3},
  pages={3380--3387},
  year={2019},
  publisher={Oxford University Press}
}

@article{yanagisawa2006first,
  title={The first confirmed Perseid lunar impact flash},
  author={Yanagisawa, Masahisa and Ohnishi, Kouji and Takamura, Yuzaburo and Masuda, Hiroshi and Sakai, Yoshihito and Ida, Miyoshi and Adachi, Makoto and Ishida, Masayuki},
  journal={Icarus},
  volume={182},
  number={2},
  pages={489--495},
  year={2006},
  publisher={Elsevier}
}

@article{madiedo2018first,
  title={The first observations to determine the temperature of a lunar impact flash and its evolution},
  author={Madiedo, Jos{\'e} M and Ortiz, Jos{\'e} L and Morales, Nicol{\'a}s},
  journal={Monthly Notices of the Royal Astronomical Society},
  volume={480},
  number={4},
  pages={5010--5016},
  year={2018},
  publisher={Oxford University Press}
}

@article{slade2010goldstone,
  title={Goldstone solar system radar observatory: Earth-based planetary mission support and unique science results},
  author={Slade, Martin A and Benner, Lance AM and Silva, Arnold},
  journal={Proceedings of the IEEE},
  volume={99},
  number={5},
  pages={757--769},
  year={2010},
  publisher={IEEE}
}

@inproceedings{ding2024china,
  title={” China Compound Eye”: Distributed Aperture Radar System for Deep Space Exploration},
  author={Ding, Zegang and Zhu, Kaiwen and Dong, Zehua and Li, Linghao and Zeng, Tao},
  booktitle={EUSAR 2024; 15th European Conference on Synthetic Aperture Radar},
  pages={1004--1007},
  year={2024},
  organization={VDE}
}

@article{savage1996interstellar,
  title={Interstellar abundances from absorption-line observations with the Hubble Space Telescope},
  author={Savage, Blair D and Sembach, Kenneth R},
  journal={Annual Review of Astronomy and Astrophysics},
  volume={34},
  number={1},
  pages={279--329},
  year={1996},
  publisher={Annual Reviews 4139 El Camino Way, PO Box 10139, Palo Alto, CA 94303-0139, USA}
}

@article{sabelhaus2004overview,
  title={An overview of the James Webb space telescope (JWST) project},
  author={Sabelhaus, Phillip A and Decker, John E},
  journal={Optical, Infrared, and Millimeter Space Telescopes},
  volume={5487},
  pages={550--563},
  year={2004},
  publisher={SPIE}
}

@article{gong2019cosmology,
  title={Cosmology from the Chinese space station optical survey (CSS-OS)},
  author={Gong, Yan and Liu, Xiangkun and Cao, Ye and Chen, Xuelei and Fan, Zuhui and Li, Ran and Li, Xiao-Dong and Li, Zhigang and Zhang, Xin and Zhan, Hu},
  journal={The Astrophysical Journal},
  volume={883},
  number={2},
  pages={203},
  year={2019},
  publisher={IOP Publishing}
}

@article{chowdhury2020imaging,
  title={Imaging infrared spectrometer onboard Chandrayaan-2 orbiter},
  author={Chowdhury, Arup Roy and Banerjee, Arup and Joshi, SR and Dutta, Moumita and Kumar, Ankush and Bhattacharya, Satadru and Rehman, Sami Ur and Bhati, Sunil and Karelia, JC and Biswas, Amiya and others},
  journal={Current Science},
  volume={118},
  number={3},
  pages={368--375},
  year={2020},
  publisher={JSTOR}
}

@inproceedings{deutsch2003mirsi,
  title={MIRSI: a mid-infrared spectrometer and imager},
  author={Deutsch, Lynne K and Hora, Joseph L and Adams, Joseph D and Kassis, Marc},
  booktitle={Instrument Design and Performance for Optical/Infrared Ground-based Telescopes},
  volume={4841},
  pages={106--116},
  year={2003},
  organization={SPIE}
}

@article{rieke2015mid,
  title={The mid-infrared instrument for the James Webb Space Telescope, VII: the MIRI detectors},
  author={Rieke, GH and Ressler, ME and Morrison, Jane E and Bergeron, L and Bouchet, Patrice and Garc{\'\i}a-Mar{\'\i}n, Macarena and Greene, TP and Regan, MW and Sukhatme, KG and Walker, Helen},
  journal={Publications of the Astronomical Society of the Pacific},
  volume={127},
  number={953},
  pages={665},
  year={2015},
  publisher={IOP Publishing}
}

@article{williams2017global,
  title={The global surface temperatures of the Moon as measured by the Diviner Lunar Radiometer Experiment},
  author={Williams, J-P and Paige, DA and Greenhagen, BT and Sefton-Nash, E},
  journal={Icarus},
  volume={283},
  pages={300--325},
  year={2017},
  publisher={Elsevier}
}

@article{paige2010diviner,
  title={Diviner lunar radiometer observations of cold traps in the Moon’s south polar region},
  author={Paige, David A and Siegler, Matthew A and Zhang, Jo Ann and Hayne, Paul O and Foote, Emily J and Bennett, Kristen A and Vasavada, Ashwin R and Greenhagen, Benjamin T and Schofield, John T and McCleese, Daniel J and others},
  journal={science},
  volume={330},
  number={6003},
  pages={479--482},
  year={2010},
  publisher={American Association for the Advancement of Science}
}

@article{robinson2010lunar,
  title={Lunar reconnaissance orbiter camera (LROC) instrument overview},
  author={Robinson, Mark S and Brylow, SM and Tschimmel, M ea and Humm, D and Lawrence, SJ and Thomas, PC and Denevi, Brett W and Bowman-Cisneros, E and Zerr, J and Ravine, MA and others},
  journal={Space science reviews},
  volume={150},
  number={1},
  pages={81--124},
  year={2010},
  publisher={Springer}
}

@misc{NASA2025YR4,
  author       = {{NASA}},
  year         = {2025},
  title        = {NASA’s Webb Observations Update Asteroid 2024 YR4’s Lunar Impact Odds},
  howpublished = {\url{https://science.nasa.gov/blogs/planetary-defense/2025/06/05/nasas-webb-observations-update-asteroid-2024-yr4s-lunar-impact-odds/}},
  note         = {Accessed: 2025-06-05}
}

@misc{JPL2025YR4,
  author       = {{NASA/JPL Small-Body Database}},
  year         = {2025},
  title        = {2024 YR4, Small-Body Database Lookup},
  howpublished = {\url{https://ssd.jpl.nasa.gov/tools/sbdb_lookup.html#/?sstr=2024%20YR4}},
  note         = {Accessed: 2025-11-25}
}

@misc{MPC2024YR4,
  author       = {{Minor Planet Center}},
  year         = {2024},
  title        = {MPEC 2024-Y140: 2024 YR4},
  howpublished = {\url{https://www.minorplanetcenter.net/mpec/K24/K24YE0.html}},
  note         = {Accessed: 2025-11-25}
}

@article{popova2011very,
  title={Very low strengths of interplanetary meteoroids and small asteroids},
  author={Popova, Olga and Borovi{\v{c}}ka, Ji{\v{r}}{\'\i} and Hartmann, William K and Spurn{\`y}, Pavel and Gnos, Edwin and Nemtchinov, Ivan and TRIGO-RODR{\'I}GUEZ, Josep M},
  journal={Meteoritics \& Planetary Science},
  volume={46},
  number={10},
  pages={1525--1550},
  year={2011},
  publisher={Wiley Online Library}
}

@article{wu2025detectability,
  title={The Detectability of Lunar-Origin Asteroids in the LSST Era},
  author={Wu, Yixuan and Jiao, Yifei and Dai, Wen-Yue and Huang, Yukun and Liu, Zihan and Cheng, Bin and Baoyin, Hexi and Li, Junfeng},
  journal={arXiv preprint arXiv:2510.23155},
  year={2025}
}

@article{baoyin2010capturing,
  title={Capturing near earth objects},
  author={Baoyin, He-Xi and Chen, Yang and Li, Jun-Feng},
  journal={Research in Astronomy and Astrophysics},
  volume={10},
  number={6},
  pages={587},
  year={2010},
  publisher={IOP Publishing}
}

@article{cheng2018collision,
  title={Collision-based understanding of the force law in granular impact dynamics},
  author={Cheng, Bin and Yu, Yang and Baoyin, Hexi},
  journal={Physical Review E},
  volume={98},
  number={1},
  pages={012901},
  year={2018},
  publisher={APS}
}

@article{devogele2025rapid,
  title={Rapid-response characterization of near-Earth asteroid 2024 YR4 during a Torino Scale 3 alert},
  author={Devog{\`e}le, Maxime and Hainaut, Olivier R and Micheli, Marco and Pravec, Petr and Cano, Juan Luis and Oca{\~n}a, Francisco and Conversi, Luca and Moskovitz, Nicholas and de Le{\'o}n, Julia and Gray, Zuri and others},
  journal={arXiv preprint arXiv:2511.09405},
  year={2025}
}

@article{lee2025trajectories,
  title={Trajectories optimization for asteroid kinetic deflection missions: Potential benefits of eccentric impacts},
  author={Lee, Kinthong and Baoyin, Hexi and Wang, Zhaokui},
  journal={Acta Astronautica},
  year={2025},
  publisher={Elsevier}
}

@article{ceplecha1998meteor,
  title={Meteor phenomena and bodies},
  author={Ceplecha, Zden{\v{e}}k and Borovi{\v{c}}ka, Ji{\v{r}}{\'I} and Elford, W Graham and ReVelle, Douglas O and Hawkes, Robert L and Porub{\v{c}}an, Vladim{\'I}r and {\v{S}}imek, Milo{\v{s}}},
  journal={Space Science Reviews},
  volume={84},
  number={3},
  pages={327--471},
  year={1998},
  publisher={Springer}
}

@article{subasinghe2018luminous,
  title={Luminous efficiency estimates of meteors. II. Application to Canadian automated meteor observatory meteor events},
  author={Subasinghe, Dilini and Campbell-Brown, Margaret},
  journal={The Astronomical Journal},
  volume={155},
  number={2},
  pages={88},
  year={2018},
  publisher={IOP Publishing}
}

@article{yu2017ejecta,
  title={Ejecta cloud from the AIDA space project kinetic impact on the secondary of a binary asteroid: I. mechanical environment and dynamical model},
  author={Yu, Yang and Michel, Patrick and Schwartz, Stephen R and Naidu, Shantanu P and Benner, Lance AM},
  journal={Icarus},
  volume={282},
  pages={313--325},
  year={2017},
  publisher={Elsevier}
}

@article{marvin1983discovery,
  title={The discovery and initial characterization of Allan Hills 81005: The first lunar meteorite},
  author={Marvin, Ursula B},
  journal={Geophysical Research Letters},
  volume={10},
  number={9},
  pages={775--778},
  year={1983},
  publisher={Wiley Online Library}
}

@article{cheng2024structural,
  title={Structural stability of China’s asteroid mission target 2016 HO3 and its possible structure},
  author={Cheng, Bin and Baoyin, Hexi},
  journal={Monthly Notices of the Royal Astronomical Society},
  volume={534},
  number={2},
  pages={1376--1393},
  year={2024},
  publisher={Oxford University Press}
}

@article{bolin2025discoveryYR4,
  title={The Discovery and Characterization of Earth-crossing Asteroid 2024 YR4},
  author={Bolin, Bryce T and Hanu{\v{s}}, Josef and Denneau, Larry and Bonamico, Roberto and Abron, Laura-May and Delbo, Marco and {\v{D}}urech, Josef and Jedicke, Robert and Alcorn, Leo Y and Cikota, Aleksandar and others},
  journal={The Astrophysical Journal Letters},
  volume={984},
  number={1},
  pages={L25},
  year={2025},
  publisher={IOP Publishing}
}
\bibliographystyle{aasjournalv7}



\end{document}